%% file: AMX_Mott.tex
\documentclass[aps,prl,twocolumn,superscriptaddress]{revtex4-1} 
\usepackage{graphicx}  
\usepackage{dcolumn}   
\usepackage{bm}        
\usepackage{amssymb}   
\begin{document}

\include{AMX_Mott_PRL}

\include{AMX_Mott_PRL_SupMat}

\end{document}

%% file: AMX_Mott_PRL.tex
\title{First-Order Insulator-to-Metal Mott Transition in the Paramagnetic \\
3D System GaTa$_4$Se$_8$}

\author{A. Camjayi}
\affiliation{Departamento de F\'{\i}sica, FCEN, UBA, and IFIBA, CONICET, Pab.~1, Ciudad Universitaria, 1428 Buenos Aires, Argentina.}
\author{C. Acha}
\affiliation{Departamento de F\'{\i}sica, FCEN, UBA, and IFIBA, CONICET, Pab.~1, Ciudad Universitaria, 1428 Buenos Aires, Argentina.}
\author{R. Weht}
\affiliation{Gerencia de Investigaci\'on y Aplicaciones, Comisi\'on Nacional de Energ\'{\i}a At\'omica (CNEA).\\ Avenida General Paz y Constituyentes, 1650 San Mart\'{\i}n, Argentina.}
\affiliation{Instituto S\'abato, Universidad Nacional de San Mart\'{\i}n-CNEA, 1650 San Mart\'{\i}n, Argentina.}
\author{M. G. Rodr\'{\i}guez}
\affiliation{Departamento de F\'{\i}sica, FCEN, UBA, and IFIBA, CONICET, Pab.~1, Ciudad Universitaria, 1428 Buenos Aires, Argentina.}
\author{B. Corraze}
\affiliation{Institut des Mat\'eriaux Jean Rouxel (IMN), Universit\'e de Nantes, CNRS,\\ 
2 rue de la Houssini\`ere, BP32229, 44322 Nantes,France}
\author{E. Janod}
\affiliation{Institut des Mat\'eriaux Jean Rouxel (IMN), Universit\'e de Nantes, CNRS,\\
2 rue de la Houssini\`ere, BP32229, 44322 Nantes,France}
\author{L. Cario}
\affiliation{Institut des Mat\'eriaux Jean Rouxel (IMN), Universit\'e de Nantes, CNRS,\\
2 rue de la Houssini\`ere, BP32229, 44322 Nantes,France}
\author{M. J. Rozenberg}
\thanks{On leave from Departamento de F\'{\i}sica, FCEN, UBA and IFIBA, CONICET}.
\affiliation{Laboratoire de Physique des Solides, CNRS-UMR8502, Universit\'e de Paris-Sud, Orsay 91405, France.}

\begin{abstract}
The nature of the Mott transition in the absence of any symmetry braking
remains a matter of debate. 
We study the correlation-driven insulator-to-metal 
transition in the prototypical 3D Mott system GaTa$_4$Se$_8$,
as a function of temperature and applied pressure.
We report novel experiments on single crystals, which demonstrate that
the transition is of first order and follows from the coexistence of two states, 
one insulating and one metallic, that we toggle with a small bias current.  
We provide support for our findings by contrasting the experimental data with
calculations that combine local density approximation with dynamical
mean-field theory, which are in very good agreement.  

\end{abstract}


\maketitle

The Mott insulator is a basic notion in modern condensed matter physics, 
at the center of strongly correlated phenomena.
In particular, the correlation-driven Mott metal-insulator transition 
takes place in systems with partially filled bands and in the absence of magnetic order
or any other type of symmetry breaking~\cite{IFT}. This transition is possibly
the simplest nontrivial strong correlation effect in a lattice system and can be considered
the electronic analogue of the familiar water-vapor transition. 
The Hubbard model is its minimal theoretical realization, and its solution
within dynamical mean-field theory (DMFT), in the 1990s, provided a detailed
scenario of the transition~\cite{review}. 
One of the key predictions is that it is of first order with
an associated region of two competing coexistent solutions: a Mott insulator and a 
correlated metal~\cite{review}. 
Their relative stability depends on the strength of
correlations, which may be physically tuned by, for instance, externally applied pressure.
One of the best candidates to exhibit the insulator-to-metal transition (IMT)  
was the compound V$_2$O$_3$, which has been continuously studied since the 1970s~\cite{mcwhan}.

While many features of the IMT in V$_2$O$_3$ do resemble the DMFT 
scenario~\cite{thomas,limelette,allen},
the comparison of the experimental data with the theory is not straightforward. 
In fact, density functional calculations
using the local density approximation (LDA) of V$_2$O$_3$ show that a multitude 
of bands cross the Fermi energy, rendering the connection to a simple Hubbard model 
physics a matter of ongoing debate~\cite{hansmann}.
For instance, the orbital degrees of freedom seem to be active
across the metal-insulator transition leading to an orbitally ordered state~\cite{paolasini}.
Moreover, the paramagnetic Mott transition does not occur for the stoichiometric compound, 
and a small chemical substitution of V by Cr is required~\cite{mcwhan}. This poses
further concerns on the validity of the single occupation constraint and on the 
eventual interplay of correlations with disorder.
Therefore, despite the extensive experimental work on V$_2$O$_3$,
which is certainly a fascinating system, the 
detailed scenario for the 3D Mott-Hubbard transition remains unclear~\cite{hansmann}. 

Here, we present a new approach to this long-standing problem 
and provide compelling experimental and theoretical
evidence of the nature of the quantum IMT in the Mott system GaTa$_4$Se$_8$ (GTS), 
which is a far simpler compound than V$_2$O$_3$.
We performed novel transport experiments on single crystals across the 
IMT as function of temperature and quasihydrostatic pressure.
We find that the transition is of first order 
due to the coexistence of two strongly correlated states, one insulating and one metallic. 
More specifically, within the transition region, where the resistivity shows nonmonotonic
behavior, we show that we may reversibly commute between the two well defined states, 
by means of a small bias current.
We further substantiate the interpretation of the experimental
data with a theoretical study of the electronic properties by means of
LDA+DMFT numerical calculations~\cite{gabiRMP}.  We find our results in very good agreement
with the present and previously reported experimental data, which provides strong support
to the DMFT Mott-Hubbard transition scenario.

GTS is a member of the family of chalcogenide compounds AM$_4$Q$_8$, 
with A=\{Ga, Ge\}, M=\{V, Nb, Ta, Mo\}, Q=\{S, Se, Te\}~\cite{prl2004,pocha2000,
pocha2005,guiot1,cario}, which were all classified as Mott 
insulators since they were expected to be metals from band structure calculations.
The importance of strong correlations was further underlined by the report 
of a pressure-driven metal-insulator crossover in GTS 
along with indications of superconductivity above 11.5 GPa
\cite{prl2004,vinhPRL} and of a doping-driven transition~\cite{dorolti}. 
Furthermore, an intriguing electric-field-driven resistive switching effect, 
both volatile~\cite{cario} and nonvolatile~\cite{vaju2008}, 
was found in many members of the family. The mechanism 
proposed for that resistive switching transition \cite{guiot,stoliar} was 
associated with Mott physics.

Unlike V$_2$O$_3$, these compounds have a rather simple structure. They are
lacunar spinels with an fcc general symmetry~\cite{barz, perrin}.
The active electrons are in the M$_4$ clusters, which are themselves forming 
an fcc sublattice. Therefore, their molecular orbitals give rise to $t_{2g}$-like conduction
bands. The electron count in GTS indicates that these bands should contain a single 
unpaired electron per Ta$_4$ cluster ~\cite{prl2004}. 
The relatively large intercluster distance results in a small hopping amplitude and, hence,
narrow partially filled conduction bands.
All these features are, in fact, borne out from band structure calculations~\cite{pocha2000, pocha2005,muller2006,epl}, which display three t$_{2g}$ bands crossing the Fermi energy.
However, as we mentioned before, all these systems turn out to be insulators.
Significantly, GTS is a paramagnetic insulator and does not order down to the lowest
measured temperatures~\cite{johrendt,pocha2005}, which is possibly due to the magnetic frustration
of the fcc arrangement of Ta$_4$ clusters. Therefore, the GTS system is relatively simple and
well suited to investigate the correlation-driven Mott transition.

\begin{figure}[!ht]
\includegraphics[width=8.0cm]{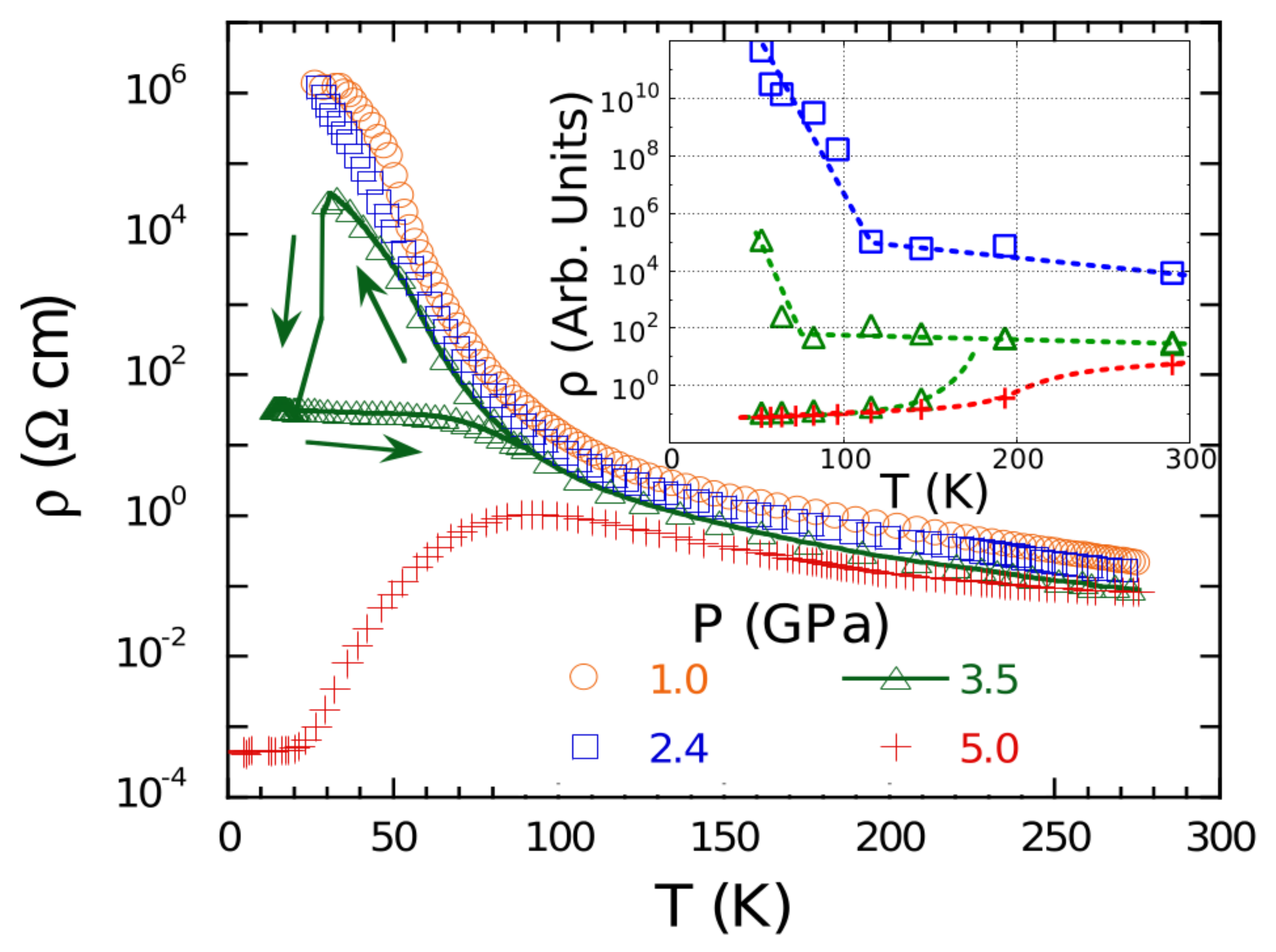}
\caption{Main panel:
Resistivity as a function of temperature (4 K $\leq T \leq$ 300 K)
for pressures from 1 to 5 GPa measured in a
GaTa$_4$Se$_8$ single crystal (sample A). 
The data show insulating behavior at low $P$ (orange circles and blue
squares), metallic behavior at high $P$ (red crosses),
and strong hysteresis
behavior at low $T \lesssim 80$ K on the metal-insulator transition region (green triangles).
Inset: LDA+DMFT results for the resistivity as a function of the
temperature. The metal (red crosses), insulator (blue squares), 
and coexistent (green triangles) solutions were obtained at $U$ =
1, 1.2, and 1.1 eV, respectively.
The lines are a guide for the eye.}
\label{fig1}
\end{figure}

In Fig.~\ref{fig1}, we show experimental data of the 
temperature dependent resistivity $\rho(T)$  of
GTS under various quasihydrostatic pressures $P$, from 1 to 5 GPa. 
The data show a huge resistivity change of
10 orders of magnitude at low temperature, as the system undergoes a dramatic 
insulator-metal transition. 
At low pressures, the system is a Mott insulator, and at high pressures, 
the state collapses to a correlated metal with anomalous resistivity. 
In fact, our measurements of $\rho(T)$ on single crystals unveil two key novel features with 
respect to previously published data obtained in polycrystalline samples~\cite{prl2004}. 
Significantly, we observe (i)
a nonmonotonic behavior in the metallic high-pressure phase ($\sim$ 5 GPa) and
(ii) a strong hysteresis effect at the critical pressure region of the Mott transition ($\sim$ 3.5 GPa). 

The nonmonotonic behavior of $\rho(T)$ is an indication of a 
strong correlation effect. In fact, it appeared in early theoretical studies 
of the single band Hubbard model within DMFT~\cite{thomas,review}. 
The experimental data show that at low $T$, the system has good metallic resistivity, which increases
with $T$. However, around 100 K,  which is merely $\sim 0.01$ eV and almost 2 orders of magnitude less
than the LDA bandwidth $\sim 0.7$ eV \cite{epl}, the resistivity suddenly flattens out
and turns semiconducting at higher temperatures.

\begin{figure}[!t]
\includegraphics[width=7.6cm]{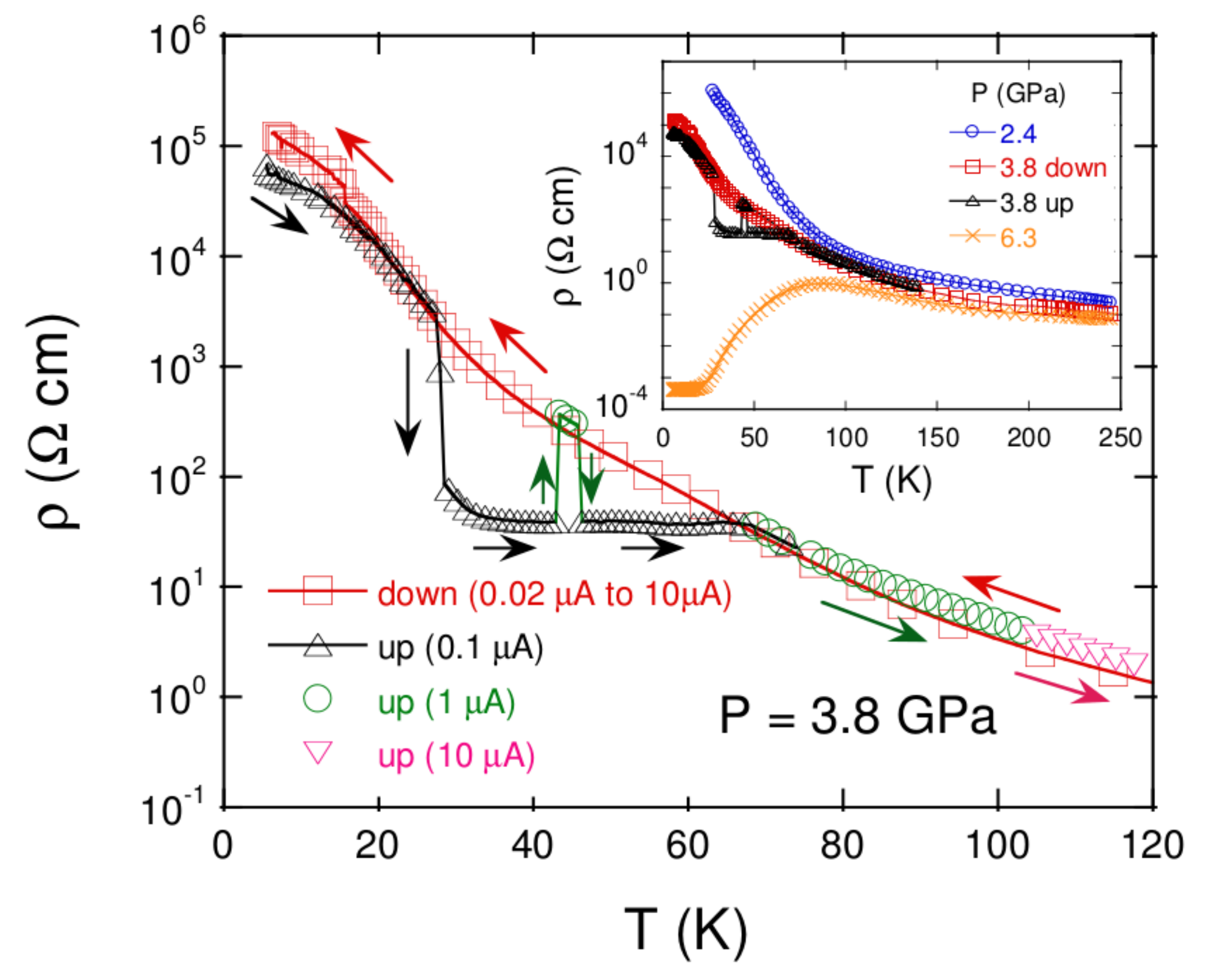}
\caption{Resistivity of a GaTa$_4$Se$_8$ single crystal (sample B)
measured within the coexistence region as a function of temperature 
at 3.8 GPa.
The commutation metal insulator in the heating process is controlled 
by a small bias current (see the text). 
Inset: Overall behavior of resistivity as a function of temperature
for different pressures (2.4 GPa for sample A and 3.8 and 6 GPa for sample B).
}
\label{fig2}
\end{figure}

Significantly, at intermediate pressures ($P$ = 3.5 GPa), within the IMT region, $\rho(T)$ shows strong hysteresis
upon cooling and heating. The pressure range of the observed hysteresis effects in different samples (see also Fig.~\ref{fig2}) 
is rather narrow and therefore easy to miss and challenging to observe. This hysteresis effect is an unambiguous indication
of the first-order character of the transition. 

We now turn the focus on the resistivity within the hysteresis region. 
One of the main findings is that the observed hysteresis is not an ordinary memory effect, 
but it is rather due to the presence of two coexistent quantum states, 
one continuously connected to the Mott insulator and the other to the correlated metal.
In Fig.~\ref{fig2} we show the strong hysteresis
effects that are observed in the transition region. 
The square (red) data points were obtained cooling down the system and the 
triangles (black and purple) and circles (green) correspond to heating.
While cooling, the applied bias current was systematically decreased, 
from 10 $\mu$A at 300 K down to 0.02 $\mu$A at 4 K, to prevent spurious Joule heating effects. 
Through the hysteresis region ($\sim$ 60 to 25 K) the bias current was 0.1 $\mu$A.  
Along the cooling process the system simply remains on the insulator state down to the lowest temperature. 
However, the activated behavior shows a smaller gap ($\sim$ 0.11 eV) with 
respect to the lower pressure data ($\sim$ 0.12 eV at 2.4 GPa and $\sim$ 0.24 eV at ambient
pressure \cite{guiot1}), indicating a further reduction of the correlation gap.
The interesting resistive bistability effect is observed upon heating.
We started to heat up applying a bias current of 0.1 $\mu$A. Initially, the data follow
essentially the same insulator-state curve obtained during the cooling process. However, at 28 K, there is
a sudden decrease of the resistance of almost 2 orders of magnitude. Upon further heating, 
$\rho(T)$ remains flat with no activated behavior, indicating the onset of a (bad) metallic state. 
At 43 K, we increased
the bias current to 1 $\mu$A to induce a small Joule heating and observed a sudden jump of the resistance, 
precisely to the insulator-state curve. We heated up the system further up, to 46 K, and it followed 
the insulator-state data.
At that point, we decreased the bias current back to 0.1 $\mu$A and observed that the resistance
immediately switched back to the metallic state curve.
Upon further heating the system remained in the metallic state. At about 70 K, the resistivity of the
heating curve merged with the cooling one, and from there on, no more hysteresis is observed.

The experiments that we just described clearly demonstrate the presence of two
coexistent states, a Mott insulator and a correlated metal, which originate the first-order 
insulator-to-metal transition. This scenario is in qualitative agreement with the early studies
of the Mott transition in the single band Hubbard model within DMFT, done in the early 1990s. 
Since then, the DMFT method has evolved from the original model
Hamiltonian studies to incorporate the realistic electronic structure of different material compounds,
which is provided by density functional methods~\cite{gabiRMP,anisimov}. 
This development enables us
to perform theoretical studies of correlation effects that are specific to each compound.
Thus, we shall provide further support to the first-order IMT scenario by means 
of an LDA+DMFT numerical study using the realistic electronic structure of GTS~\cite{epl}.
Details of the numerical methods are provided in the Supplemental Material~\cite{supmat}.

An important issue of the LDA+DMFT method is the determination of the magnitude of the
correlation parameter $U$, originated in the local Coulomb interactions.
Its {\it ab initio} determination remains a challenging problem.
Here, we shall use a physical argument to estimate its value.
For simplicity, we shall also assume a single $U$ value for all interorbital 
repulsion parameters and we neglect the Hund's interaction.
From experiments we know that GTS is, in fact, a Mott insulator; hence, the parameter
$U$ must be large enough to destroy the LDA metallic state.
However, a second experimental fact is that under the effect of external pressure, GTS
undergoes an insulator-to-metal transition. External pressure produces an increase of the
bandwidth $W$ of about 0.023 eV/GPa  \cite{vinhPRL}.
It thus decreases the strength of the correlation effects $U/W$ (since $U$ is rather independent of pressure)
and leads to the onset of the IMT in GTS.
Therefore,  from these observations, we may conclude that the value of interaction $U$ should be such
that places the system on the insulator side and near the IMT. Therefore, our strategy will be
to first estimate the value of interaction $U$ that destroys the LDA metal and then to study the behavior of
the system in the neighborhood of the transition to compare with the experiments. 
For simplicity, we shall use $U$ as a parameter and keep the LDA band structure fixed,
since the conclusions of our study do not depend on this choice (see 
the Supplemental Material~\cite{supmat} for details).

\begin{figure}[!hb]
\includegraphics[width=8.0cm]{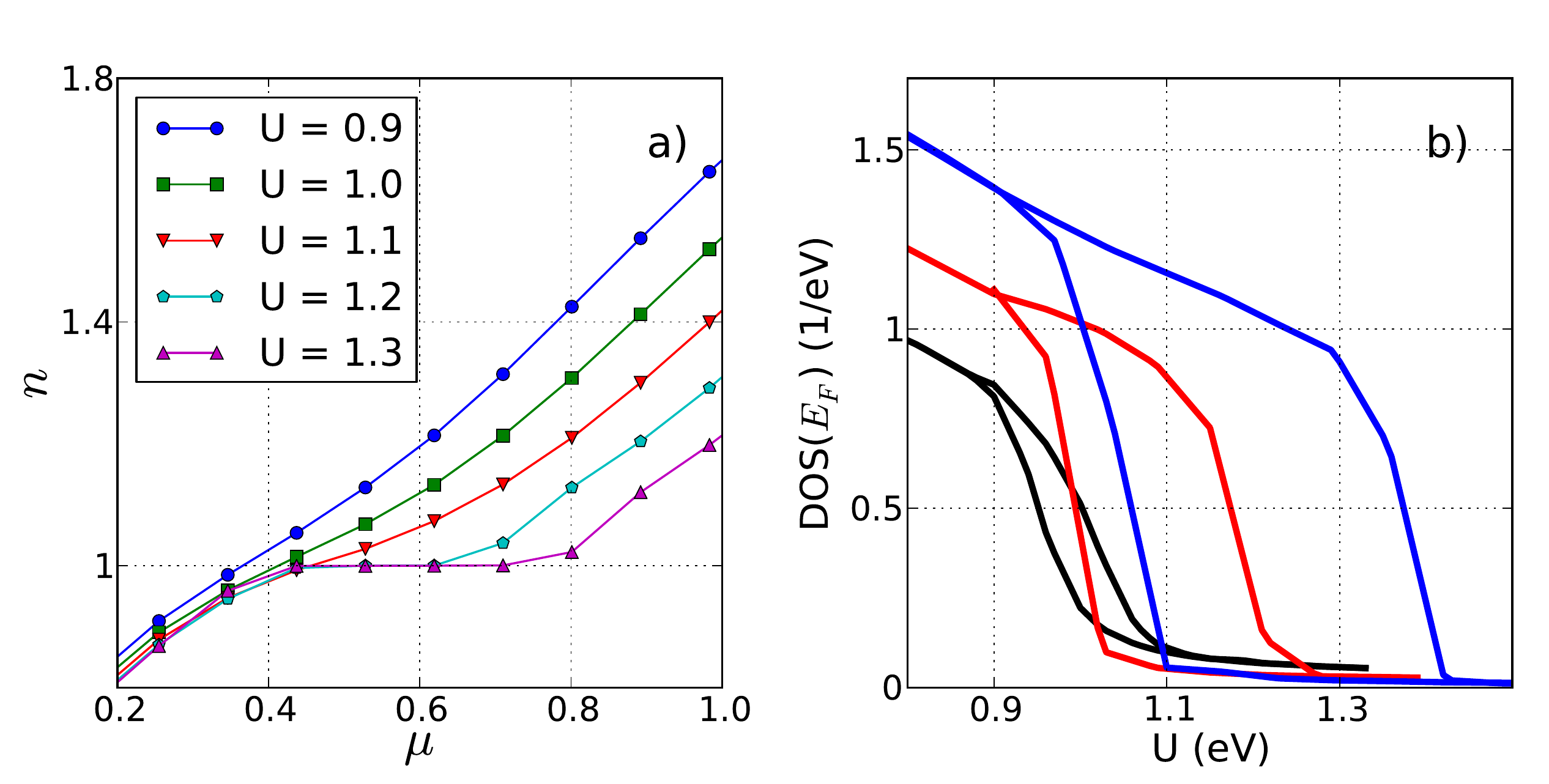}
\caption{(a) LDA+DMFT results of the occupation of the
t$_{2g}$ bands of GTS as a function of the chemical potential $\mu$ and the 
interaction $U$. The data show that the Mott gap opens at $n=1$ for
$U\approx 1.2$ eV. (b) Density of states
at the Fermi energy as a function of $U$ for $n = 1$ 
at $T$ = 58, 116 and 232~K (top to bottom), 
indicating the coexistence of solutions. 
}
\label{fig3}
\end{figure}

In Fig.~\ref{fig3}(a), we show the LDA+DMFT results for the electron occupation 
versus chemical potential 
as a function of the local Coulomb repulsion $U$.
We observe that for strong enough interaction, the $n \mbox{ vs. }\mu$ curves develop a
plateau at $n=1$, which corresponds to the filling of GTS (i.e., one electron per Ta$_4$ tetrahedron).
The plateau denote the destruction of the metal and the onset of the incompressible Mott insulator
state \cite{rozenbergPRB}.  
From the data, we estimate that the GTS system at ambient pressure has a value of  $U \approx 1.2$ eV. 

We now explore the neighborhood of the IMT to search for coexistent solutions and hysteresis.
To numerically search for coexistent solutions, we simply ``follow'' one solution, say, the insulating one, and slowly 
lower the value of $U$ (i.e., the $U$/$W$ ratio) recomputing the converged solution at every step. Eventually, the
solution has a sudden change to a qualitatively different state, i.e. the metallic solution. Then, we follow
the same procedure for the metallic solution, now increasing the value of $U$, until we observe a sudden change. 
The region of $U$ values where two converged solutions are found determines the coexistent region. 
In Fig.~\ref{fig3}(b)
we show the hysteresis of the density of states (DOS) at the Fermi energy, which clearly 
distinguishes the insulator from the metal state.
The coexistence region extends in a small window of $U/W$ values, 
narrows as the temperature is increased, and eventually
disappears above $T \approx$ 230 K. 
This is consistent with our experimental results, which showed two coexistent states within
a narrow range of pressures, below $T \approx$ 100 K. This agreement is remarkable, since 
it should be kept in mind that this temperature is quite a small energy scale, almost 2 orders of 
magnitude smaller than the LDA bandwidth $W$ and the interaction $U$, which emerges
from a many-body effect.
Within LDA+DMFT, we may also compute the resistivity $\rho(T)$ of the system. The results
are shown in the inset of Fig.~\ref{fig1}, which show a good agreement with
the experimental results. The numerical data capture the basic qualitative features,
namely, the relatively weak variation of $\rho$ at higher $T$, the many orders
of magnitude of difference between the metallic and insulating resistivities at low $T$,
and the nonmonotonic behavior of the correlated metal state.

\begin{figure}
\includegraphics[clip,width=8.0cm]{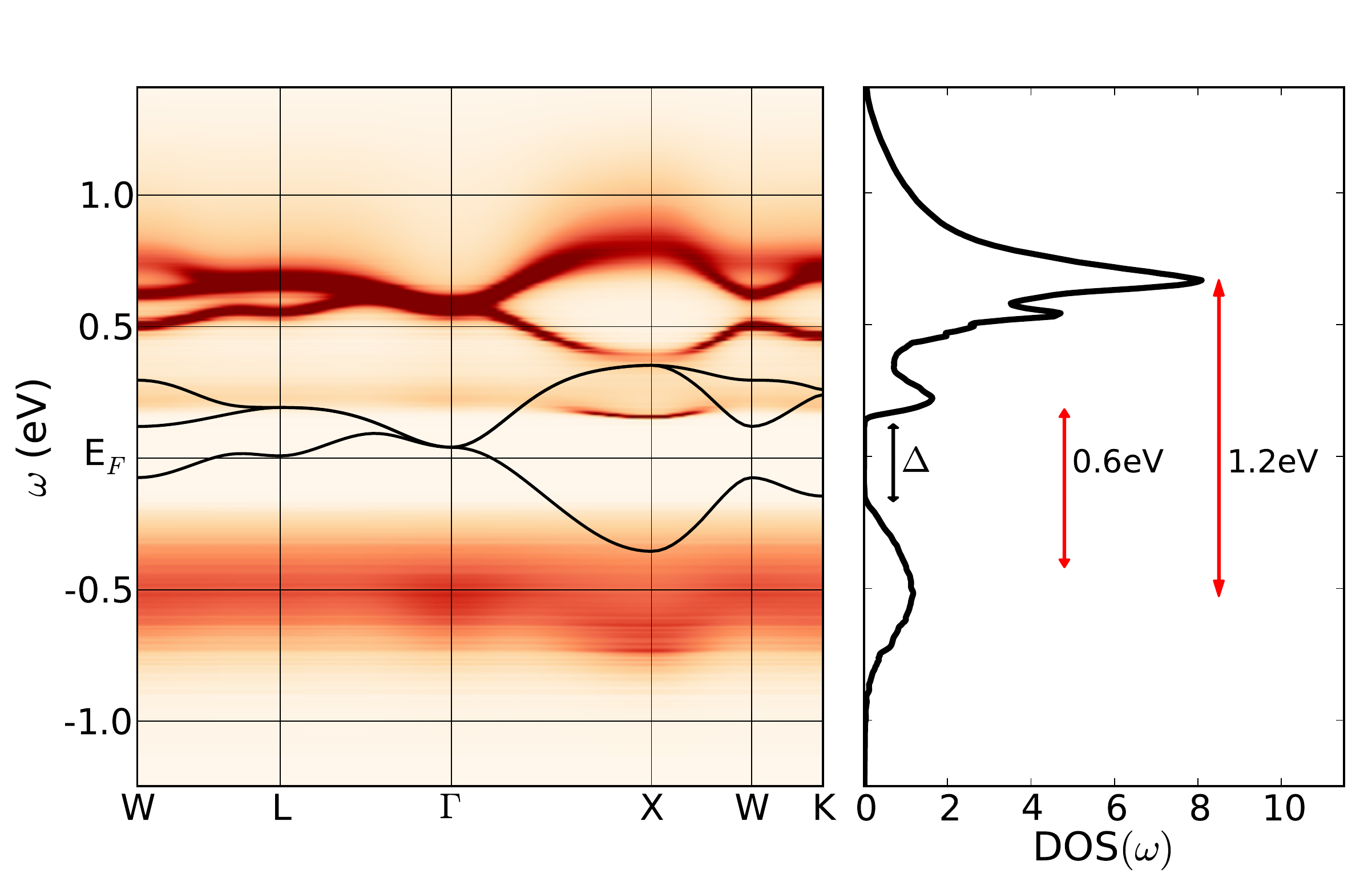}
\caption{LDA+DMFT calculation of the Mott insulator state.
Left panel: Intensity plot of the electronic structure of the
t$_{2g}$ bands of GTS obtained from calculations for $U$ = 1.2 eV and $T$ = 58 K.
Note within the upper Hubbard band at the $X$ point the presence of a lower 
energy feature (see the Supp. Material~\cite{supmat}).
The strength of the occupied part of the spectra ($\omega <$ 0) has been enhanced
by a factor of 4 to enable a better visualization.
For reference, we also include the standard LDA calculation in solid black lines \cite{epl}.
Right panel: The total local DOS($\omega$). The (red) arrows indicate
the main contributions that should be observed in optical conductivity.
The (black) arrow indicates the small transport gap $\Delta$, consistent
with the value measured in resistivity experiments~\cite{vinhPRL}.
}
\label{fig4}
\end{figure}

Finally, we turn to the spectral functions, which reveal some unexpected features
in regard to recent experiments \cite{vinhPRL}.
The interacting band structure, as expected, displays a strong renormalization
with respect to the LDA calculations. The results are shown in Fig.~\ref{fig4}.
For reference, we also display in the figure the LDA bands. 
The Mott insulator has lower and upper Hubbard bands, which are split
by energy $\sim U$. They are very incoherent but still bear resemblance to the LDA dispersion.
In the figure, we also show the local DOS($\omega$).
Interestingly, our results enable us to interpret and understand the apparent inconsistency between
our presently estimated value of $U$ = 1.2 eV and the smaller one $\sim 0.55$ eV reported in a recent
optical conductivity study of GTS \cite{vinhPRL}.
Our numerical calculation reveals the existence of a secondary structure within the upper Hubbard band,
a narrow subband at $\omega \sim$ 0.2 eV.
This feature should lead to the presence of two distinct contributions in the optical conductivity spectrum,
as indicated in Fig.~\ref{fig4} a more prominent one of magnitude $\omega \sim$ 1.2 eV and a smaller 
one at $\sim$ 0.6 eV. This latter one is consistent with the value of the excitation reported in 
Ref.~\onlinecite{vinhPRL}. 
The origin of this contribution can be traced to a multiorbital effect, due to the strong orbital polarization of the LDA
bands at the $X$, $Y$, and $Z$ points of the Brillouin Zone (see the Supplemental Material~\cite{supmat}). The detailed study of this point and the calculation of the
optical response are beyond the scope of the present study.
Finally, we should also mention that the small gap in the DOS($\omega$) of magnitude $\Delta \approx 0.30$ eV
is also consistent with the value 0.241 eV reported from the activated behavior
of the resistivity at ambient pressure \cite{guiot1}. 

To conclude, we presented a combined experimental and theoretical study of the Mott-Hubbard 
insulator-to-metal transition in the strongly correlated system GaTa$_4$Se$_8$.
Our resistivity experiments under quasihydrostatic pressure on GTS single crystals
enable us to uncover the first-order character of the correlation-driven
paramagnetic 3D Mott transition.
Moreover, we achieved control of a resistive bistability at the transition, 
providing validation of its physical nature, as was qualitatively predicted by DMFT more 
than 20 years ago.
We have provided further and decisive support by means of GTS specific LDA+DMFT 
calculations, which are in good agreement with the present and recently reported experiments.

\begin{acknowledgments}
We thank V.~Ta Phuoc, V. Guiot, and P. Stoliar for useful discussions.
We acknowledge support from the French Agence Nationale de la Recherche (ANR-09-Blan-0154-01)
and (R.W. and C.A.) from CONICET and ANPCyT (PIP 114-201101-00376, PICT-2012-0609, and PIP 112-201101-00536).
\end{acknowledgments}

{
\onecolumngrid
\begin{center}
\newpage
\end{center}
\twocolumngrid
}

%% file: AMX_Mott_PRL_SupMat.tex
\makeatletter 
\renewcommand{\thefigure}{S\@arabic\c@figure}
\makeatother
\setcounter{figure}{0}

\setcounter{page}{1}
\renewcommand{\thepage}{SM-\arabic{page}}

{
\onecolumngrid
\begin{center}
\large\bf
{Supplemental Material of 
``First-order insulator to metal Mott-transition in the 
paramagnetic 3D system GaTa$_4$Se$_8$''}
\end{center}
\hspace{0in}
\hspace{0.5in}
\twocolumngrid
}

\section{Experimental details}

The high pressure was applied using
a quasi-hydrostatic experimental setup,
corresponding to a Bridgman
configuration with WC anvils, where pyrophillite is used as a gasket
and steatite as the pressure medium that favors quasi-hydrostatic
conditions. Pressure inside the cell was determined by measuring the
resistive superconducting transition of Pb. The pressure gradient
was estimated from the width of this transition and corresponds to a
5-10\% of the applied pressure. A standard 4 terminal DC technique
was used to measure resistivity under high pressure at different
temperatures. The typical size of the samples was of a fraction of a mm.
Electrical contacts were made using 25 $\mu$m diameter
Pt wires pressed to the sample's surface by the pressure setup. A
well calibrated carbon-glass thermometer thermally anchored to the
anvils ensures a determination of sample's temperature with an
uncertainty lower than 0.5 K for the whole temperature range
studied.

\section{Numerical calculation details}

\subsection{LDA}
In the LDA calculation we use the Wien2K code~\cite{SM-wien2k}, that is an implementation of
the full-potential linearized-augmented plane wave method (FP-LAPW)~\cite{SM-lapw}.
As we are not interested in total energies nor magnetic states, we adopt the simplest 
local density approximation to represent the exchange-correlation potential~\cite{SM-lda}.
Results using the generalized gradient approximation are completely equivalent
since band structures are, in general, insensitive to this choice.
Even thought within the DFT schemes the eigenvalues and eigenfunctions formally have no direct physical meaning, 
it is, nevertheless, broadly accepted that they provide a good approximation for the quasi-particle 
energies and band-structure.  Thus, here we shall adopt them as 
the reference for the calculation of the localized
Wannier orbital basis \cite{SM-epl}. 
These Wannier orbitals represent the molecular orbitals arising from the Ta$_4$ clusters, which are well defined 
as the distances between the four metal atoms in the cluster are significantly 
shorter than the inter-cluster ones.
We restrict the energy window to the three $t_{2g}$ bands that
cross the Fermi energy. We note that these three bands are quite well separated from the rest of the
band manifold, which is very advantageous in order to successfully obtain an accurate 
local orbital basis representation with short ranged hopping amplitudes.
The localized Wannier orbital basis is computed following the procedure described by
Marzari, Vanderbilt and coworkers in a series of papers~\cite{SM-MLWF} and implemented in
the code Wannier90~\cite{SM-W90}. As the interface between the two programs we use 
the code wien2wannier~\cite{SM-w2w}.

\begin{figure}[!ht]
\includegraphics[clip,width=8cm]{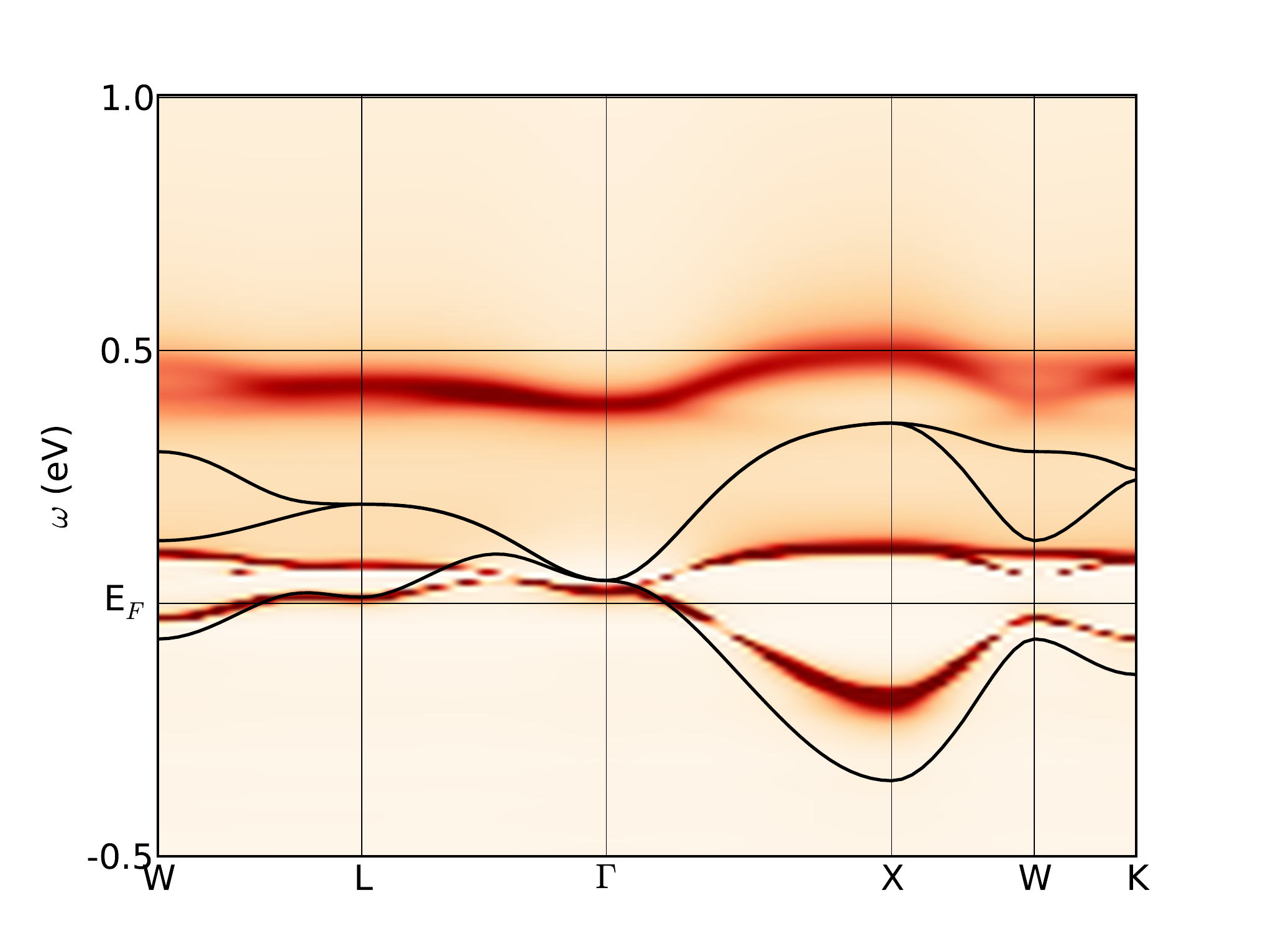}
\caption{Intensity plot of the electronic structure of the
t$_{2g}$ bands of GTS obtained from LDA+DMFT calculations for $U$ = 1.2 eV and $T$ = 58 K
within the coexistence region. The data correspond to the metallic solution that
coexists with the insulator one, shown in Fig.~4 of the main text.
For reference, we also include the standard LDA calculation in solid black lines \cite{SM-epl}.
}
\label{figS1}
\end{figure}

\subsection{DMFT}
The DMFT calculation is done using a continuous-time quantum Monte Carlo (CTQMC)
impurity solver \cite{SM-review_gull}, based on the hybridization expansion~\cite{SM-haule}.
In order to perform reliable analytic continuation of the imaginary axis data
a large number of Monte Carlo steps has to be adopted to minimize statistical
errors.
Special care has been taken to obtain well converged solutions in the coexistence region,
which is complicated by critical slowing down and enhanced statistical fluctuations.
Due to the general cubic symmetry of the system, the double-counting problem is absent.

The correlation effects to the LDA bandstructure is obtained via a DMFT self-energy, which is calculated 
from an auxiliary quantum-impurity problem. The numerical calculations are done with a 
multi-orbital CTQMC method, which allows us to reach the relevant temperature regime. 
To explore experimentally the IMT, external pressure was applied, which affects the (bare) bandwidth $W$ and,
hence, the strength of electronic correlations $U/W$. In principle, one could simulate this by recomputing
the LDA bandstructure at the compressed atomic positions. This methodologically demanding
and full empirical information is not available, thus,
we shall proceed in a different simpler manner. 
From previous LDA calculation under pressure \cite{SM-vinhPRL}, it is possible to see that 
the bandstructure is modified by a simple scale factor in the range of pressures considered experimentally, 
without changing qualitatively the dispersion relation. Since we choose the same value of $U$ for all inter-orbital
interactions, we  shall keep the ambient pressure LDA electronic structure fixed and
use $U$ as the free parameter to change the $U/W$ ratio.

\subsection{LDA+DMFT metallic solution}
For completeness, we show in Fig.~\ref{figS1} the LDA+DMFT results for the
bandstructure of the correlated metallic solution at $U$ = 1.2 eV and
$T$ = 58 K. 
The electronic structure near the Fermi energy shows a narrowing with respect
to the LDA bands, which reflects the mass renormalization of the correlated
metallic state. In addition, at higher energies one finds spectral weight associated
to the Hubbard bands. The lower Hubbard band cannot be observed in the intensity
plot due to its relatively weak spectral strength.

\subsection{Resistivity}
The calculation of the resistivity is done following the 
definition \cite{SM-oudovenko}: $\rho = \frac{k_B T}{e^2}\frac{1}{A_0}$ with
\begin{equation}
A_0 = \lim_{\omega \to 0} \left[ \frac{\hbar}{\omega_n \beta}\int_0^\beta \mathrm{d}\tau e^{i\omega_n\tau}
\langle \langle j^x(\tau) j^x(0) \rangle \rangle \right]_{\omega_n \to \omega + i 0^+}
\end{equation} 
We evaluate the correlation function $\langle \langle j^x(\tau) j^x(0) \rangle \rangle$ 
within DMFT, in the dipole approximation and ignoring vertex corrections.

\section{Orbital polarization in k-space}

The LDA bandstructure in the maximally localized Wannier orbital representation is
the starting point for the LDA+DMFT calculation. We adopt the construction made
in Ref.~\onlinecite{SM-epl}. In Fig.~S2, we reproduce 
the bandstructure of the three t$_{2g}$ fatbands,
which indicate the orbital character across the Brillouin zone.

\begin{figure}
\includegraphics[clip,width=8cm]{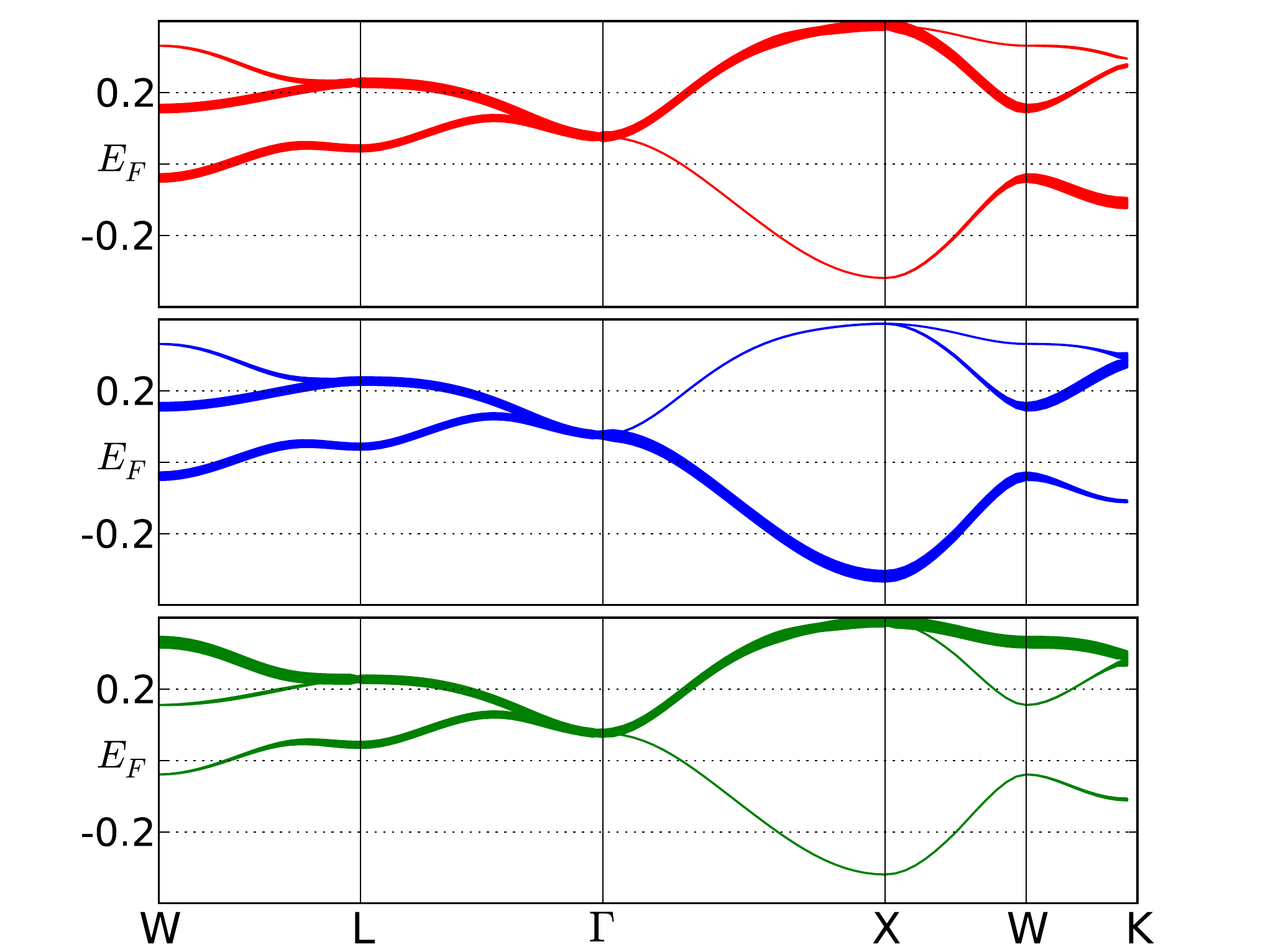}
\caption{LDA calculated ``Fat bands'' of GTS (bottom). Red, blue and green 
respectively correspond to $d_{xy}$,  $d_{yz}$ and $d_{zx}$ characters.
}
\label{figS2}
\end{figure}

As expected, at the $\Gamma$-point the three bands are degenerate, as dictated by
the cubic lattice symmetry. We also observe that from the $\Gamma$- to the $L$-point
the orbital character of the three bands is almost degenerate. Interestingly, however,
the situation is dramatically different from  the $\Gamma$- to the $X$-point. In this case
we observe that at the $X$-point the character of the band beneath the Fermi level 
is dominated by the $yz$ orbital. Similarly, by symmetry, we have $zx$ at the $Y$-point 
and $xy$ at the $Z$-point. Hence, while the system is degenerate at the $\Gamma$-point,
there is a strong orbital polarization at the $X$, $Y$ and $Z$-points.
This feature, which we shall explore in detail elsewhere is responsible for the small
band that is observed at low (positive) energies at the $X$ point of the upper Hubbard 
band, in Fig.~4 of the main text.